\documentclass[aps,prl,preprint,superscriptaddress]{revtex4}

\usepackage{graphicx}
\usepackage{dcolumn}
\usepackage{bm}
\usepackage{amsmath}
\usepackage{amssymb}
\usepackage{latexsym}
\usepackage{epsfig}
\usepackage{amsbsy}
\usepackage{array}
\usepackage{amssymb}
\usepackage{setspace}
\usepackage{bm}

\def\sint{\ifmmode{- \!\!\!\!\!\! \int}
    \else{\hbox{$- \!\!\!\! \int \ $}}\fi}

\begin{document}


\title{Is silicene stable in air?-First principles study of oxygen adsorption and dissociation on silicene$^{*}$}

\author{G. Liu$^{\rm 1,  2}$, X. L. Lei$^{\rm 1,  2}$, M. S. Wu$^{\rm 1,  2}$, B. Xu,}

\author{C. Y. Ouyang \footnote{ electronic mail:cyouyang@jxnu.edu.cn. }}

\affiliation{College  of Physics  and  Communication Electronics,
Jiangxi Normal University, Nanchang, Jiangxi, 330022, P. R. China}

\affiliation{Key Laboratory of Photoelectronic and Telecommunication
of Jiangxi Province, Nanchang, Jiangxi, 330022, P. R. China}


\date{\today}

\begin{abstract}
The oxygen adsorption and dissociation on pristine silicene surface
are studied by use of first-principles in this letter. The oxygen
adsorption and dissociation on pristine silicene surface are studied
by use of first-principles in this letter. It is found that the
pristine silicene is not stable in air because the oxygen molecule
can be easily adsorbed and dissociated into two O atoms without
overcoming any energy barrier on pristine silicene surface. In
addition, dissociated oxygen atoms are relatively difficult to
migrate on or desorbed from pristine silicene surface, leading to
poor mobility of oxygen atom. As a result, silicene would be changed
into Si-O compounds in air. The work will be helpful to reveal the
detail of the interaction between oxygen molecules and pristine
silicene surface, especially helpful to understand the stability of
silicene in air.
\end{abstract}


\pacs{Valid PACS appear here}

\maketitle

\section{Introduction}  
Slicene, the silicon-based counterpart of graphene, has promoted
from theoretical predictions to experimental observations in last
few years\cite{1,2,3,4,5}. Theoretically, density functional theory
(DFT) calculations on silicene show that ${\pi}$ and $\pi^{*}$ bonds
linearly cross at the Fermi level, reflecting the semi-metallic or
zero-gap semiconducting character of silicene\cite{6}. In addition,
most of the other known features of silicene resemble those of
graphene. Therefore, due to its unique structure and electronic
properties of graphene-like two-dimension (2D) sheet, silicene has
the potential to provide a new future for the electronics industry.
Especially, it could be expected to offer an easily implemented
alternative for the enhancement of the performance and scalability
of the present silicon-based electronics. Interestingly, even before
the synthesis of graphene, first-principles local-density
approximation calculations predicted that a buckled honeycomb
structure of Si could exit\cite{7,8}, in contrast with the planar
honeycomb structure of graphene. Furthermore, on the basis of the
first-principles calculations of structure optimization, phonon
modes, and finite temperature molecular dynamics, it is reported
that although the planar and high-buckled structure of Si is
unstable, the low-buckled (LB) honeycomb structure can be stable
with an equilibrium buckling of $\Delta$=0.46 {\AA}\cite{9}.

Experimentally, Nakano \emph{et al.} has reported the synthesis of
silicene via the chemical exfoliation of CaSi$_{2}$\cite{10}.
Recently, the possible growth of the silicene sheet on Ag (111) has
been reported\cite{11}. For example, Patrick Vogt \emph{et al.}
report that they have provided compelling evidence for the synthesis
of epitaxial silicene sheets on Ag (111) substrates though the
combination of scanning tunneling microscopy and angular-resolved
photoemission spectroscopy in conjunction with calculations based on
density functional theory\cite{12}. But, it is worthy to be
mentioned that, from all of the experiments of silicene synthesis,
the Ag substrates may play a catalyst role in the formation and
stabilization of the silicene sheet. More importantly, silicene
sheet, particularly, only can be successfully synthesized in
ultrahigh vacuum conditions and the experimental conditions for
producing this silicon structure are quite strict. Thus, there is an
urgent question to silicene sheet for its future applications: If
the substrate and ultrahigh vacuum conditions are got rid of, is the
pristine silicene stable in air? Further, if it is not truer, the
previous theoretical studies on silicene may be no longer
comprehensive.

To investigate the stability of silicene in air and reveal the
details of its possible changes, oxygen adsorption and dissociation
on silicene is studied by use of first-principles calculations in
this letter. It is found that pristine silicene is not stable in air
because the oxygen molecule can be adsorbed easily on the silicene
sheet surface without overcoming any energy barrier, resulting in
dissociation of oxygen molecule on silicene surface. Furthermore,
dissociated oxygen atoms are correspondingly difficult to migrate on
the silicene surface or desorbed from the silicene, leading to poor
mobility of oxygen atom, especially, the Si-O compounds.

\section{method}  
All calculations are performed by using the VASP (Vienna \emph{ab}
initio simulation package) within the projector augmented-wave (PAW)
approach\cite{13}. The ground state of the electronic structure is
described within density functional theory (DFT) using the
generalized gradient approximation (GGA) with PW91 exchange
correlation functional\cite{14}. The energy cutoff for expansion of
wave functions and potentials is 550 eV. The single layer silicene
sheet is modeled with a 6$\times$6$\times$¡Á1 unit cell containing
72 Si atoms, which is separated with a 15 {\AA} vacuum layer in the
z-axis direction. Monkhorst-Pack special k-point method\cite{15}is
used with a grid of 1$\times$1$\times$1. The entire systems are
relaxed by conjugate gradient method until the force on each atom is
less than 0.05 eV/{\AA}. To optimize the O$_{2}$ molecule
dissociation path, the climbing image nudged elastic band (CINEB)
method\cite{16} is used in the present study.

\section{Results and Discussions}
\subsection{adsorption of single O atom on silicene}
To study the adsorption and dissociation of O$_{2}$ molecule on
pristine silicene surface, a Low-buckled silicene sheet with the
super cell 6$\times$6$\times$1 has been firstly optimized and a
lattice constant of \emph{a }= 23.15 {\AA} and ¦¤=0.46 {\AA} are
obtained (corresponding a = 3.86 {\AA} in unit cell ), in well
agreement with the previous report\cite{17}. After that, single O
atom is put on the pristine silicene surface with four typical
adsorption sites. They are the center site of the Si$_{6}$ ring, the
bridge site of two nearest Si atoms, the top site of a Si atom, the
top site of a neighboring Si atom at same layer, marked with No. 1,
No. 2, No. 3, and No. 4, respectively, as shown in Fig. 1(a). Sites
No. 3 and No. 4 are different due to the buckling of the silicene
sheet. We calculated the adsorption energy (\emph{E}$_{ad}$), which
is defined as:
$$
E_{ad}=-[E(Si_{64}O)-E(Si_{64})-1/2\times E(O_{2})]
$$
where the \emph{E}(\emph{Si}$_{64}$\emph{O}) and
\emph{E}(\emph{Si}$_{64}$) are the total energy of the silicene
supercell model with and without adsorption of one O atom,
respectively.\emph{ E }(\emph{O}$_{2}$) is the total energy of
O$_{2}$ molecular in vacuum. The calculated adsorption energy
\emph{E}$_{ad}$ and adsorption distance \emph{D }(defined as the
distance from the adsorbate to the substrate plane, namely, the
difference between the \emph{z}-axis coordinate of the adsorbate and
the average of \emph{z}-axis coordinate of all surface atoms) for
the adsorption of an O atom on different sites of silicene sheet are
listed in Table 1.

It is found that the strongest adsorption energy of the adsorbed O
atom on silicene is observed on the bridge sites (Site No. 2) of
Si$_{6}$ ring with the adsorption energy of 2.395 eV and the
adsorption distance 1.544 {\AA}, reflecting the most stable
adsorption site is the bridge site. The corresponding Si-O bond
length of neighboring Si atoms is 1.714 {\AA} as well as another is
1.734 {\AA}. The minor difference of two Si-O bong lengths
originates from covalent interaction between O atom and neighboring
Si atoms, resulting in one Si atom moving up after adsorption, as
shown in Fig. 1(b).

\subsection{migration of single O atom on silicene}

It is well known that O atoms are ready to react with other foreign
atoms when O$_{2}$ molecule is dissociated into O atoms and O atom
is adsorbed at the bridge sites of two nearest Si atoms. Before the
reaction happens, O atoms need to diffuse on the silicene surface to
meet other foreign atoms. Therefore, the migration of O atom on the
silicene surface is also necessary to be considered. In order to
find the optimized migration pathways, the climbing nudged elastic
band (CNEB) method is used. After calculations, it is found that
there are two migration pathways for O atom to move. One is 2-3-2,
another is 2-4-2, respectively, as shown in Fig. 2 (a), where No. 2
and No. 4 represent different top site along two pathways. Next, the
energy profile along the optimized migration pathway for single O
atom migrated on the silicene surface is obtained, as shown in Fig.
2 (b). Moreover, it is found that the energy barrier is 1.183 eV as
well as another is 1.048 eV, which are almost equal. Compared with
the energy barrier (0.72 eV) of O atom migration on graphene that
has been reported by H. J. Yan \emph{et al.}\cite{18}, it shows that
the O atom migration will be more difficult on silicene surface,
resulting in the poor mobility of O atom under room temperature.
Therefore, it implies that O atom is not easy to move from one
bridge site to another bridge site on silicene surface.

\subsection{adsorption and dissociation of O$_{2}$ molecule on silicene}

Owing to the similar structure and electronic properties between
silicene and graphene, it is worthwhile to mention that O$_{2}$
molecule can be stable at the graphene sheet. It has been reported
that when one O$_{2}$ molecule put on the graphene sheet with
different orientations, it prefers to stay parallel to the graphene
basin plane and locate in the middle of a C$_{6}$ ring with an
adsorption height of 2.82 {\AA}. It implies that O$_{2}$ molecule
can be adsorbed physically on the graphene surface\cite{18}.
Moreover, dissociation of O$_{2}$ molecule on graphene is an
endothermal process, and the energy barrier of the dissociation
reaction is very high (2.39 eV). Those data shows that dissociation
of O$_{2}$ at the graphene sheet is physically not favorable.
Combined with other previous reports\cite{19,20} , we can conclude
that the pristine graphene is stable in air. However, we demonstrate
that the O$_{2}$ at the silicene sheet is in a completely different
manner.

To investigate the stability of silicene in air, the oxygen
adsorption and dissociation on silicene must be studied. In
addition, to imitate the actual condition, the influence of the
concentration of O$_{2}$ molecule for adsorption and dissociation
can not be ignored. Considering the probable influence of the
concentration of O$_{2}$ molecule for adsorption and dissociation on
silicene surface, we test the pristine silicene under different
concentration of O$_{2}$ molecule by use of unit cell
1$\times$1$\times$1, super cell 2$\times$2$\times$1, and super cell
3$\times$3$\times$1, respectively. Firstly, the unit cell
(1$\times$1$\times$1) is used to adsorb two O atoms. The
concentration of O$_{2}$ molecule is 100\% in terms of unit cell
(composed of two Si atoms). It is found that, locating on the top of
the Si atoms, one O atom is up as well as another is down. The
corresponding Si-O bond length is all 1.554 {\AA}. Subsequently,
super cell 2$\times$2$\times$1, and 3$\times$3$\times$1 are used to
test the adsorption and dissociation of O$_{2}$ molecule with the
initial distance 2.7 {\AA} both vertical and parallel to the
silicene basin plane. After calculations, the optimized results show
that whatever the concentration of O$_{2}$ is, O$_{2}$ molecule can
be easily adsorbed and dissociated into two O atoms. It implies that
the concentration of O$_{2}$ molecule has no influence to its
adsorption on silicene surface. Thus, we still adopt super cell
6$\times$6$\times$1 mentioned above to accomplish the next work.

 To study the adsorption and dissociation of O$_{2}$ molecule
on pristine silicene, O$_{2}$ molecule is put above the silicene
surface with different orientations (both parallel and vertical to
the silicene basin plane) and distance. There are two different
types and every type includes three different cases. One is O$_{2}$
molecule put above the silicene surface vertically with three
different initial distances, 1.48 {\AA}, 2.70 {\AA} and 4.23 {\AA},
respectively. Another is O$_{2}$ molecule put above the silicene
surface parallel with three different initial distances, 1.50 {\AA},
2.70 {\AA} and 4.15 {\AA}, respectively. It is found that the
O$_{2}$ molecule can be easily adsorbed and dissociated into two O
atoms on bridge site and top site on silicene surface except for the
case with initial distance 4.23 {\AA} vertical to substrate plane,
which corresponding adsorption energy is about 0 eV, reflecting
hardly any interaction between oxygen and silicene surface. Instead,
although the O-O bond is not broken in the case with initial
distance 4.15 {\AA} parallel to substrate plane, but the O-O bond
length of 1.52 {\AA} in contrast with the free O-O bond length 1.23
{\AA} shows that the interaction of two O atoms is very weak. On the
other hand, from all of the calculated adsorption energy, the 4.916
eV with initial distance 1.50 {\AA} vertical to silicene surface is
highest, indicating the corresponding structure is the most stable
structure where the O$_{2}$ molecule is dissociated into two O atoms
on bridge site and top site in same Si$_{6}$ ring. Therefore, the
case with initial distance 2.70 {\AA} parallel to the silicene plane
is used to illustrate the detail of the oxygen adsorption and
dissociation on silicene.

Compared with dissociation reaction energy 2.39 eV of graphene in
air, when we put oxygen molecule above the silicene surface basin
plane with initial distance 2.70 {\AA}, it is surprisingly found
that the oxygen molecule can be easily adsorbed and dissociated into
two O atoms on silicene surface without overcoming any energy
barrier. That is to say, the silicene is not stable in air. To
visually illustrate the detail of O$_{2}$ molecule adsorption and
dissociation on silicene surface, the evolution from free O$_{2}$
molecule to dissociated O atom is shown in Fig. 3. When the free
oxygen molecule is put above the silicene surface with initial
distance 2.7 {\AA}, the corresponding O-O bond length is 1.235
{\AA}, as shown in Fig. 3(a). As the optimized process continues,
the distance between the oxygen molecule and silicene surface basin
decreases, leading to one O atom adsorbed on top site on silicene
surface with the total energy of -353.919 eV, as shown in Fig. 3(b).
It is worthwhile to note that the orientation of oxygen has changed
from vertical to almost parallel to silicene surface, while the
corresponding O-O bond length and the Si-O bond length are 1.563
{\AA} and 1.720 {\AA}, respectively. It indicates that the
interaction between oxygen and silicene surface becomes stronger and
the O-O bond strength becomes weakened. Subsequently, the
dissociation of oxygen molecule happens, resulting in one O atom
adsorbed on bridge site of neighboring Si atom with Si-O bond length
of 1.740 {\AA} and 1.711 {\AA} but another still adsorbed on top
site with Si-O bond length of 1.565 {\AA}, as shown in Fig. 3 (c).
The corresponding total energy decreases to -357.798 eV. Finally,
the optimized structure shows that the oxygen molecule dissociates
into two O atoms. One O atom locates on top site where the Si-O bond
length is 1.570 {\AA}, while another O atom adsorbs on bridge site
where the Si-O bong lengths are 1.711 {\AA} and 1.732 {\AA},
respectively, as shown in Fig. 3(d). The corresponding total energy
evolution can be seen in Fig. 3(e), in which the majorization
structures are marked with (a), (b), (c), and (d). The abrupt turn
of total energy changes from -353.919 eV(marked with (b)) to
-357.798 eV (marked with (c)) exactly reflects the oxygen
dissociation on silicene.

\section{Summary}  

In summary, the oxygen adsorption and dissociation on pristine
silicene surface are studied by use of first-principles in this
letter. It is found that the pristine silicene is not stable in air
because the oxygen molecule can be easily adsorbed and dissociated
into two O atoms without overcoming any energy barrier on pristine
silicene. In addition, dissociated oxygen atoms are relatively
difficult to migrate on the silicene surface or desorbed from
pristine silicene surface, leading to poor mobility of oxygen atom
and the Si-O compounds. The work will be helpful to understand the
stability of silicene in air and will widen the application of
silicene. Moreover, it implies that some previous works on pristine
silicene are not comprehensive.

Thanks to the support of NSFC under Grant Nos. 11064004, 11234013
and 11264014. Thanks to the support of the Open Foundation of Key
Laboratory of Photoelectronic and Telecommunication of Jiangxi
Province (Grant No. 2013009).


\newpage

\begin{table}[h]
\caption{Calculated adsorption energy \emph{E}\emph{$_{ad}$} and
adsorption distance \emph{D} for the adsorption of an O atom on
different sites of silicene sheet.}
\begin{center}
\begin{tabular}{llllllllllll} 

\hline\hline Sites & 1(center)& 2(bridge)& 3(top1) &(4(top2) \\

\hline \emph{D}({\AA})\ \ & 0 & 1.544 & 2.406 & 2.237 \\

\hline \\\emph{E}$_{ad}$(eV)\ \ & -0.694 & 2.395 & 1.379 & 1.241 \\

\hline \hline
\end{tabular}
\end{center} \label{tab:phase}
\end{table}

\newpage

A list of figures

\begin{enumerate}

\item (Color online) (a) Schematic views of four adsorption site for single O atom on pristine silicene surface and (b) top and side views of the bridge sites.
\label{fig:Graph1}

\item  (Color online) (a) Top view of the optimized migration pathway of an O atom migration along the pathway from site No. 2 to site No. 3 then to site No. 2 and its energy profile along the pathway on the silicene surface. (b) Top view of the optimized migration pathway of an O atom migration along the pathway from site No. 2 to site No. 4 then to site No. 2 and its energy profile along the pathway on the silicene surface. (Red color and yellow color represent O atoms and Si atoms, respectively)  \label{fig:Graph2}

\item (Color online) The diagrammatic sketch of oxygen adsorption and dissociation on silicene.
\label{fig:Graph3}

\end{enumerate}

\newpage

\begin{figure}
\centering
\begin{minipage}[b]{0.5\textwidth}
\centering
\includegraphics[width=3in]{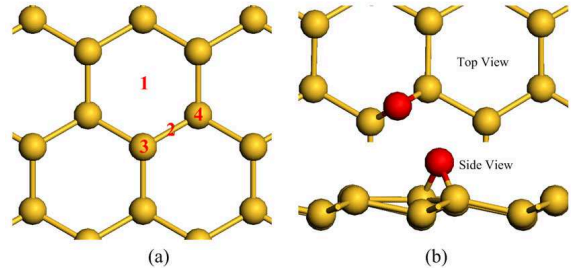}
\end{minipage}
\caption{(Color online) (a) Schematic views of four adsorption site
for single O atom on pristine silicene surface and (b) top and side
views of the bridge sites. \label{fig:Graph1}}
\end{figure}

\newpage

\begin{figure}
\centering
\begin{minipage}[b]{0.5\textwidth}
\centering
\includegraphics[width=4in]{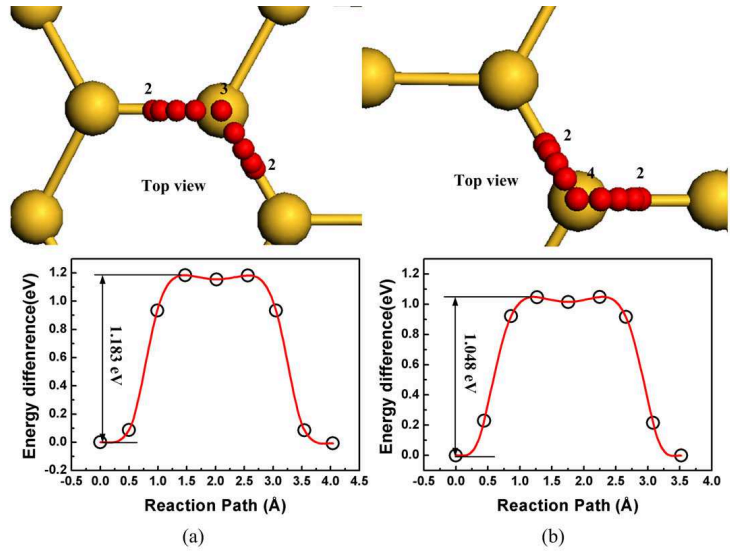}
\end{minipage}
\caption{(Color online) (a) Top view of the optimized migration
pathway of an O atom migration along the pathway from site No. 2 to
site No. 3 then to site No. 2 and its energy profile along the
pathway on the silicene surface. (b) Top view of the optimized
migration pathway of an O atom migration along the pathway from site
No. 2 to site No. 4 then to site No. 2 and its energy profile along
the pathway on the silicene surface. (Red color and yellow color
represent O atoms and Si atoms, respectively) \label{fig:Graph2}}
\end{figure}

\newpage

\begin{figure}
\centering
\begin{minipage}[b]{0.5\textwidth}
\centering
\includegraphics[width=4in]{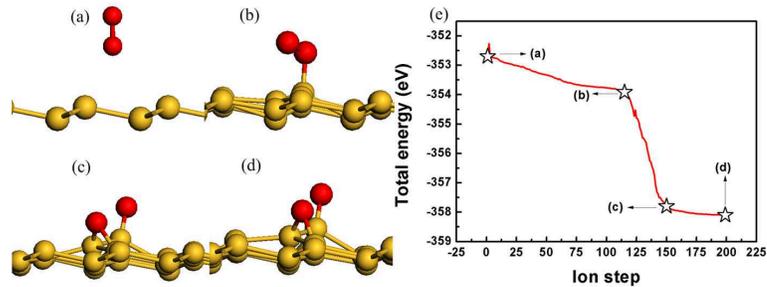}
\end{minipage}
\caption{(Color online) The diagrammatic sketch of oxygen adsorption
and dissociation on silicene. \label{fig:Graph3}}
\end{figure}

\newpage

\end{document}